\documentclass[letterpaper, 10 pt, conference]{ieeeconf}  

\IEEEoverridecommandlockouts                              
\usepackage{graphics} 
\usepackage{amsmath} 
\usepackage{amssymb}  
\usepackage{braket}
\usepackage{soul}
\usepackage{xcolor}
\usepackage{hyperref}
\usepackage{adjustbox}
\usepackage{stmaryrd}

\hypersetup{
    colorlinks=true,
    linkcolor=blue,
    }

\title{\LARGE \bf
Scaling QAOA: transferring optimal adiabatic schedules from small-scale to large-scale variational circuits}

\author{Ugo \textsc{Nzongani}${}^{1, 2}$\thanks{${}^{1}$Aix-Marseille Université, Université de Toulon, CNRS, LIS, 13288 Marseille, France. Contact: \href{mailto:ugo.nzongani@lis-lab.fr}{ugo.nzongani@lis-lab.fr}}, Dylan \textsc{Laplace Mermoud}${}^{2, 3}$\thanks{${}^{2}$UMA, ENSTA, Institut Polytechnique de Paris, 91120 Palaiseau, France}\thanks{${}^{3}$CEDRIC, Conservatoire National des Arts et Métiers, 75003 Paris, France}, Arthur \textsc{Braida}${}^{4}$\thanks{${}^{4}$Université Paris Cité, CNRS, IRIF, Paris, France} 
}

\begin{document}

\maketitle
\thispagestyle{empty}
\pagestyle{empty}

\begin{abstract}
The Quantum Approximate Optimization Algorithm (QAOA) is a leading approach for combinatorial optimization on near-term quantum devices, yet its scalability is limited by the difficulty of optimizing \(2p\) variational parameters for a large number \(p\) of layers. Recent empirical studies indicate that optimal QAOA angles exhibit concentration and transferability across problem sizes. Leveraging this observation, we propose a schedule-learning framework that transfers spectral-gap-informed adiabatic control strategies from small-scale instances to larger systems. 

Our method extracts the spectral gap profile of small problems and constructs a continuous schedule governed by \(\partial_t s = \kappa g^q(s)\), where \(g(s)\) is the instantaneous gap and \((\kappa, q)\) are global hyperparameters. Discretizing this schedule yields closed-form expressions for all QAOA angles, reducing the classical optimization task from \(2p\) parameters to only \(2\), independent of circuit depth. This drastic parameter compression mitigates classical optimization overhead and reduces sensitivity to barren plateau phenomena.

Numerical simulations on random QUBO and 3-regular MaxCut instances demonstrate that the learnt schedules transfer effectively to larger systems while achieving competitive approximation ratios. Our results suggest that gap-informed schedule transfers provide a scalable and parameter-efficient strategy for QAOA.
\end{abstract}



\section{Introduction}
Quantum computing promises to solve specific classes of problems with exponentiel or polynomial speedups over classical counterparts, most notably in cryptography \cite{shor1999polynomial}, database searching \cite{grover1996fast}, and quantum simulation \cite{daley2022practical}. However, identifying other domains where quantum computers yield a practical computational advantage remains an active area of investigation. A primary candidate is combinatorial optimization, where problems are naturally mapped onto qubit Hamiltonians whose solution spaces grow exponentially with the number of variables \cite{abbas2024challenges}.

Historically, quantum optimization methods have been deeply rooted in adiabatic processes \cite{Albash_2018}. In Adiabatic Quantum Computing, the system slowly interpolates between a trivial mixing Hamiltonian and a problem Hamiltonian whose ground state encodes the optimal solution \cite{lucas2014ising}. The trajectory of this interpolation is governed by an annealing schedule. The standard adiabatic criterion guarantees finding the final ground state. It scales the evolution speed inversely proportional to the square of the minimum spectral gap, which often yields exponential runtimes for hard instances.~\cite{AKR_2010}. 

In the Noisy Intermediate-Scale Quantum (NISQ) era \cite{preskill2018quantum}, the dominant paradigm for optimization on gate-based devices is the Variational Quantum Algorithm (VQA) \cite{cerezo2021variational, 10.1145/3569095, LaplaceMermoud2026variationalquantum}. The leading candidate in this class is the Quantum Approximate Optimization Algorithm (QAOA) \cite{farhi2014quantumapproximateoptimizationalgorithm}. QAOA can be viewed as a trotterized, variational approximation of the adiabatic evolution, alternating between unitary evolutions generated by the mixer and problem Hamiltonians. While QAOA has shown numerical promise, demonstrating scaling advantages for the classically intractable LABS problem \cite{Shaydulin_2024}, it introduces a new challenge: parameter optimization. The optimization landscape is frequently plagued by barren plateaus, where gradients vanish exponentially, making classical training of the circuit angles intractable for deep circuits \cite{larocca2025barren}.

Interestingly, the dynamics of QAOA also form a subclass of Continuous-Time Quantum Walks (CTQW) \cite{farhi1998quantum}, representing a parameterized walk on a graph of feasible solutions under a potential field induced by the problem Hamiltonian. Recent works emphasize CTQWs as strong candidates for optimization \cite{callison2019finding,banks2024continuous}, where the schedule corresponds to the hopping rate of the walker rather than a strict adiabatic interpolation. Various methods have been proposed to approximate these schedules to guide the walker toward high-quality solutions \cite{schulz2024guided,nzongani2025sampled}.

Whether viewed through the lens of CTQWs or adiabatic evolution, determining the optimal schedule remains a bottleneck. Recent studies on QAOA parameter concentration demonstrate that angles optimized on small instances can be transferred to larger problems \cite{pelofske2024scaling,pelofske2025evaluating}. In this work, we propose a heuristic that leverages this transferability by learning the adiabatic schedule directly from the spectral gap scaling of small instances. By mapping the continuous adiabatic path to the discrete QAOA circuit, we reduce the optimization complexity from $2p$ parameters (where $p$ is the number of layers in the trotterization) to just 2 global hyperparameters, significantly mitigating the classical optimization overhead.

\section{Background}

\subsection{QUBO and Ising Formulation}
The most extensively studied class of combinatorial problems in the quantum computing community is Quadratic Unconstrained Binary Optimization (QUBO). A QUBO instance is defined by the objective function 
\[
f_Q(x)=\sum_i Q_{ii}x_i+\sum_{i<j}Q_{ij}x_ix_j, 
\]
where $Q\in\mathbb{R}^{n\times n}$ is a coefficient matrix and the goal is to find the bitstring $x^*\in\{0,1\}^n$ that minimizes $f_Q$. 

By mapping binary variables to spin variables via $x_i=(1-z_i)/2$ with $z_i\in\{+1,-1\}$, the QUBO problem is equivalent to finding the ground state of an Ising spin-glass Hamiltonian:
\begin{equation}\label{eq:ising_hamiltonian}
    H_1 = \sum_i h_i\sigma^z_i + \sum_{i<j}J_{ij}\sigma^z_i\sigma^z_j,
\end{equation}
where the local fields and couplings are given by $h_i=-Q_{ii}/2-\sum_{j\neq i}Q_{ij}/4$ and $J_{ij}=Q_{ij}/4$, respectively.

\subsection{Adiabatic Quantum Computing and Schedules}
Adiabatic Quantum Computing relies on the adiabatic theorem, which guarantees that a quantum system remains in its instantaneous ground state if the Hamiltonian evolution is sufficiently slow. The process interpolates between an initial Hamiltonian $H_0$, whose ground state is easy to prepare, and the problem Hamiltonian $H_P$ (Eq. \eqref{eq:ising_hamiltonian}). A standard choice is the transverse field Hamiltonian $H_0=-\sum_i\sigma^x_i$, with the ground state $\ket{+}^{\otimes n}=2^{-n/2}\sum_{x}\ket{x}$. The time-dependent Hamiltonian is given by:
\begin{equation}\label{eq:adiabatic_hamiltonian}
    H(s) = (1-s) H_0 + s H_1,
\end{equation}
where $s(t)\colon [0, T] \to [0, 1]$ is the annealing schedule, parameterizing the evolution from $t=0$ to the total time $T$.

The efficiency of this algorithm is dictated by the system's spectral properties. Let $E_0(s)$ and $E_1(s)$ denote the instantaneous ground and first excited state energies of $H(s)$, respectively. The instantaneous spectral gap is defined as:
\begin{equation*}
   g(s)=E_1(s)-E_0(s).
\end{equation*}
From recent theoretical work on the adiabatic theorem \cite{jansen2007bounds, Braida2025unstructured}, the schedule $s(t)$ that optimizes the total running time of the evolution to end up close to the final ground state, should follow the shape of the instantaneous gap. Specifically, the optimal schedule $s(t)$ adapts the evolution speed $ds/dt$ such that it is maximized where the gap is large and minimized where the gap is small (the bottleneck). Analytically, this optimal velocity scales as $ds/dt \propto g^q(s)$ (for $q \in (1,2)$).

\subsection{Transferability of Spectral Profiles}
In practice, computing the instantaneous gap $g(s)$ to tailor the optimal schedule is computationally intractable for large $n$, as it requires the diagonalization of an exponentially large matrix. However, for many classes of random optimization problems, the structural features of the spectral gap, such as the location of the minimum gap, often exhibit self-similar behavior across different system sizes due to the concentration of measure~\cite{10.1098/rsta.2021.0417, irsigler2022quantum}. 

In this work, we exploit this property to bypass the computational bottleneck. We propose to “learn” the optimal schedule profile from tractable, small-scale instances (e.g. $n=10$) by explicitly calculating their spectral gaps. We then extrapolate this learned mean schedule to solve larger target instances (up to $n=20$).

\begin{figure}
    \centering
    \includegraphics[scale=0.4]{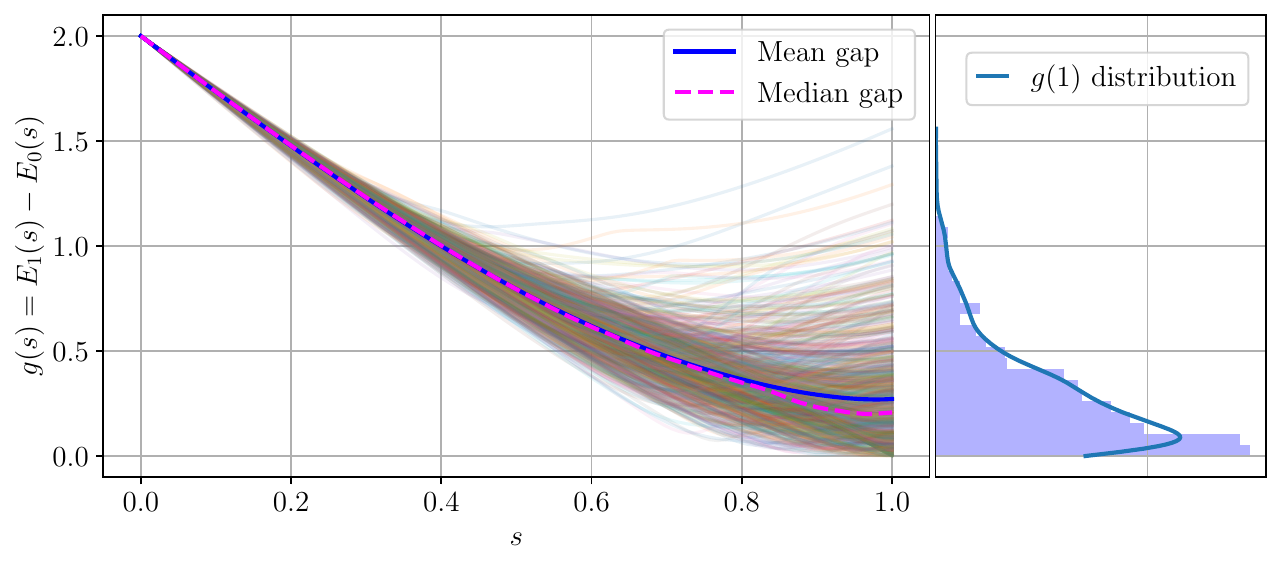}
    \caption{Sampled instantaneous gaps for $n=10$ qubits on 1000 QUBOs with coefficients randomly drawn in [-1,1] (left) and the final gap distribution (right).}
    \label{fig:sampled_gaps}
\end{figure}

\section{Schedule Learning Framework}

Our heuristic strategy aims to bypass the high-dimensional classical optimization landscape of QAOA by leveraging insights from optimal adiabatic control. In this section, we first outline the continuous framework used to learn the target schedule shape from the spectral gap. We then demonstrate how to map this continuous schedule onto the discrete variational parameters of a quantum circuit. Notably, this mapping reduces the dimension of the classical optimization problem to just two hyperparameters ($\kappa$ and $q$), independent of the number of layers in the trotterization $p$.

\subsection{Sampling the Gap to Learn the Schedule}

The efficiency of adiabatic evolution is governed by the system's spectral gap. To learn the optimal schedule, we first perform a “learning phase” on small QUBO instances. In this phase, we treat the adiabatic parameter $s$ simply as a normalized reaction coordinate $s \in [0, 1]$ (equivalent to a linear schedule $s(t) = t/T$ where $T$ is the total time) to diagonalize the instantaneous Hamiltonian $H(s)$ and map the spectral gap profile $g(s) = E_1(s) - E_0(s)$.

With the gap profile $g(s)$ known, we construct an optimized non-linear time evolution $s(t)$ designed to satisfy the adiabatic condition. We define the evolution rate to be proportional to a power of the gap:
\begin{equation}\label{eq:gap_derivative}
    \frac{ds}{dt} = \kappa g^q(s),
\end{equation}
where $\kappa$ is a proportionality constant controlling the total time, and $q$ is a power-law exponent (typically $1 \le q \le 2$ in theoretical bounds, but treated here as a learnable hyperparameter).

To derive the dynamics in terms of the path parameter $s$, we apply the chain rule to the time-dependent Schrödinger equation (with $\hbar=1$):
\begin{equation}
    i \frac{d}{dt}\ket{\psi(t)} \hspace{-1pt}=\hspace{-1pt} H(t)\ket{\psi(t)} \hspace{-3pt}\implies\hspace{-2pt} i \frac{ds}{dt} \frac{d}{ds}\ket{\psi(s)} \hspace{-1pt}=\hspace{-1pt} H(s)\ket{\psi(s)}.
\end{equation}
Substituting our ansatz for the schedule velocity $ds/dt$, we obtain the effective Schrödinger equation in the $s$-domain:
\begin{equation}\label{eq:schrodinger_s}
    i \kappa g^q(s) \frac{d}{ds}\ket{\psi(s)} = H(s)\ket{\psi(s)}.
\end{equation}
The formal solution to this equation generates the unitary evolution operator $U_{tot}$ acting on the initial state $\ket{\psi(0)} = \ket{+}^{\otimes n}$:
\begin{equation}
    \ket{\psi(s=1)} = \mathcal{T}\exp\left(-i\int_0^1 \frac{H(\sigma)}{\kappa g^q(\sigma)} d\sigma\right)\ket{\psi(0)},
\end{equation}
where $\mathcal{T}$ denotes the time-ordering operator.

\subsection{Quantum Circuit Derivation}

To implement this evolution on a gate-based quantum computer, we approximate the continuous integral by discretizing the path $s \in [0,1]$ into $p$ steps of width $\delta s = 1/p$. The integral is approximated as a product of exponentials (Riemann sum):
\begin{equation}
    U_{tot} \simeq \prod_{k=1}^p \exp\left(-i \frac{\delta s}{\kappa g^q(s_k)} H(s_k) \right),
\end{equation}
where $s_k = k \cdot \delta s$. Recall that $H(s_k) = (1-s_k)H_0 + s_k H_1$. Since $H_0$ and $H_1$ do not commute, we apply a first-order Trotter-Suzuki decomposition $e^{-i(A+B)} \approx e^{-iA}e^{-iB}$ to separate the mixing and problem Hamiltonians:
\begin{equation}\label{eq:trotterized_circuit}
    U_{tot} \approx \prod_{k=1}^p e^{-i \beta_k H_0} e^{-i \gamma_k H_1}=U_p(\kappa, q).
\end{equation}
By matching the terms in the exponents, we derive the explicit formulas for the variational angles at layer $k$:
\begin{equation}\label{eq:angles_mapping}
    \gamma_k = \frac{s_k \delta s}{\kappa g^q(s_k)}, \qquad \beta_k = \frac{(1-s_k) \delta s}{\kappa g^q(s_k)}.
\end{equation}
These equations constitute the core of our mapping: the angles are no longer $2p$ independent variables, but are fully determined by the learned gap function $g(s)$ and the two global hyperparameters $\kappa$ and $q$.

\subsection{Circuit implementation and rescaling.}
The unitary operators in Eq.~\eqref{eq:trotterized_circuit} are implemented using standard gate sets. The mixer Hamiltonian $H_0 = -\sum \sigma^x_i$ and the problem Hamiltonian $H_1$ (Eq.~\eqref{eq:ising_hamiltonian}) decompose as:
\begin{align}
    e^{-i\beta_k H_0} &= \prod_{l=1}^n R_X^{(l)}(-2\beta_k), \\
    e^{-i\gamma_k H_1} &= \left(\prod_{l=1}^{n}R_Z^{(l)} \left( 2\gamma_k \tilde{h}_l \right) \right) \left(\prod_{i<j} R_{ZZ}^{(i,j)} \left( 2\gamma_k \tilde{J}_{ij} \right) \right),
\end{align}
where $R_{ZZ}^{(i,j)}(\theta) = CX_{ij} R_Z^{(j)}(\theta) CX_{ij}$.

Finally, to ensure consistent behavior across random instances with varying energy scales, we normalize the problem coefficients during the learning and implementation phases. The rescaled Ising coefficients $\tilde{h}$ and $\tilde{J}$ are such that their range of values is the same as that of the learning part. This preprocessing step ensures that the sampled gaps remain transferable regardless of the specific numerical range of the QUBO coefficients.

\section{Numerical simulations}

\subsection{Variational framework}

We parametrize the circuit of Eq. \eqref{eq:trotterized_circuit} with $\vec{\theta}=(\kappa,q)$ and use a classical optimizer to find the best hyperparameters $\vec{\theta}$ that minimize the energy expectation value 
\[
E(\psi(\vec{\theta}))=\bra{\psi(\vec{\theta})}H_1\ket{\psi(\vec{\theta})} \text{ with } \ket{\psi(\vec{\theta})} = U_p(\vec{\theta})\ket{+}^{\otimes n}.
\]

We compare our heuristic with vanilla QAOA \cite{farhi2014quantumapproximateoptimizationalgorithm}, whose parameterized circuit reads:
\begin{equation}
    \ket{\psi(\vec{\gamma},\vec{\beta})}=\left(\prod_{k=1}^p e^{-i\beta_k H_0}e^{-i\gamma_k H_1}\right)\ket{+}^{\otimes n},
\end{equation}
where $\vec{\gamma}=(\gamma_1,\dots,\gamma_k)$ and $\vec{\beta}=(\beta_1,\dots,\beta_k)$ are the $2p$ tunable parameters obtained through classical optimization.

We use the approximation ratio as a metric of performance, which measures the distance between the resulting quantum state and the ground state of the problem Hamiltonian:
\begin{equation}
    r(\psi(\vec{\theta}))=\frac{E_{\max}-E(\psi(\vec{\theta}))}{E_{\max}-E_{\min}},
\end{equation}
where $E_{\max}$ and $E_{\min}$ are the maximum and minimum eigenvalues of the problem Hamiltonian $H_1$, where the eigenvectors associated to $E_{\min}$ encode the optimal solutions to the optimization problem.

\subsection{Experimental settings}

We perform the learning part by computing the spectral gap $g(s)$ on 1000 QUBO problems of size $n=10$ with random coefficients uniformly drawn in $[-1,1]$ as shown in Fig.~\ref{fig:sampled_gaps}. Therefore, for QUBO problems with coefficients drawn in $[Q_{\min},Q_{\max}]$ with $(Q_{\min},Q_{\max})\neq (\pm 1,\pm 1)$, we rescale the Ising coefficients as $\tilde{h}_{i}=\alpha h_i$ and $\tilde{J}_{ij}=\alpha J_{ij}$ with $\alpha=2/(Q_{\max}-Q_{\min})$.

\begin{figure}
    \centering
    \includegraphics[scale=0.25]{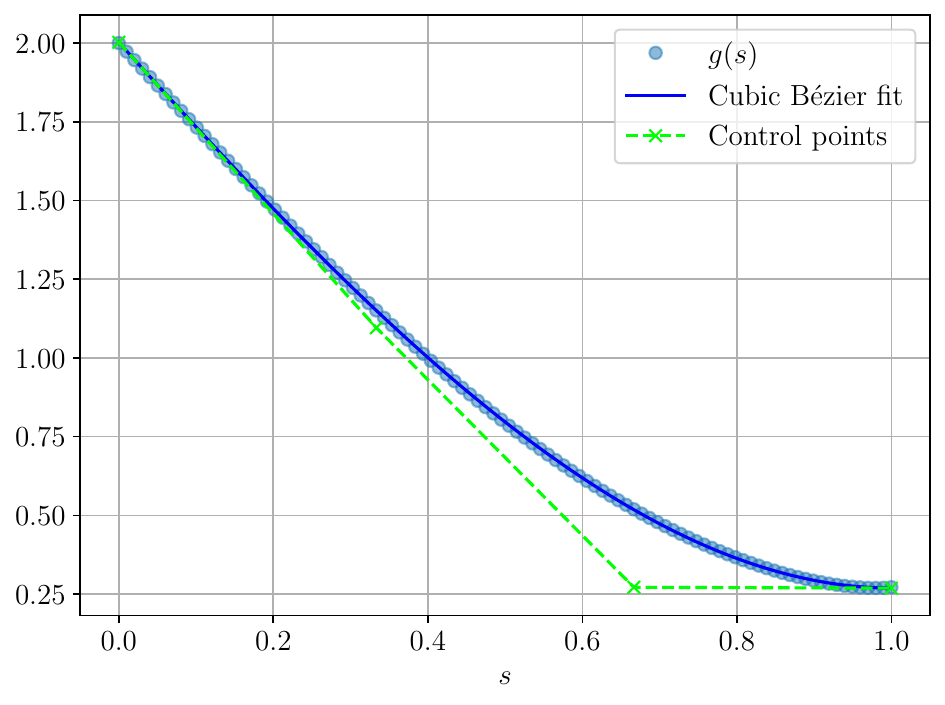}
    \includegraphics[scale=0.25]{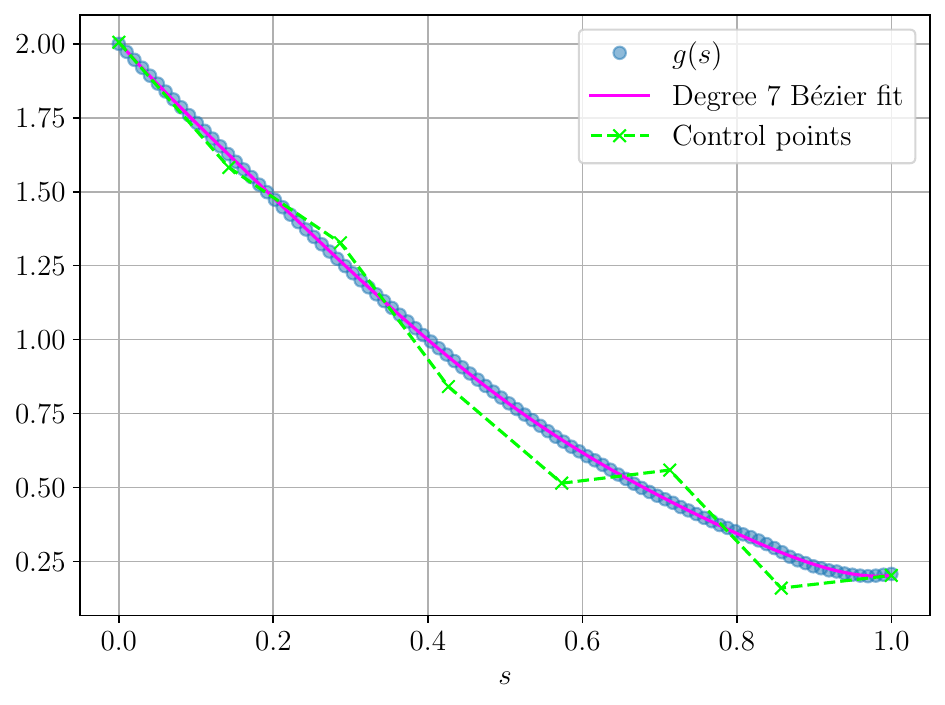}
    \caption{Bézier fitting of the mean (degree 3) and median (degree 7) sampled gaps obtained in the learning phase.}
    \label{fig:bezier}
\end{figure}

We use the mean and median sampled gaps as candidates for the gap $g$ of Eq. \eqref{eq:gap_derivative} used to obtain the schedule derivative, and we use Bézier curves to derive their analytical expressions:
\begin{equation*}
	B(t) = \sum_{i=0}^n \begin{pmatrix} n \\ i\end{pmatrix}(1-t)^{n-i}t^iP_i,
\end{equation*}
with $0\leq t\leq 1$ and $\{P_i=(x_i,y_i)\}$ the set of control points. We respectively fit the mean and median sampled gaps with Bézier curves of degrees 3 and 7 as shown in Fig. \ref{fig:bezier}, and we identify the gap as the $y$-axis components of the curves\footnote{Thus, we have $g(f)=(1-f)^3y_0+3(1-f)^2fy_1+3(1-f)f^2y_2+f^3y_3$ in the case of cubic Bézier fitting.}.

We compare our heuristic with vanilla QAOA on 3-regular unweighted MaxCut, 3-regular weighted MaxCut and random QUBO problems. For each problem, we generate 100 different instances of $n=20$ qubits and simulate the circuits with $p\in \llbracket 1,10 \rrbracket$ layers of problem and mixer Hamiltonians in Qiskit \cite{qiskit2024}. The classical optimization of the 2 (our heuristic) and $2p$ (QAOA) hyperparameters is done with 200 iterations of COBYLA.

\subsection{MaxCut}

Given a graph $G=(V,E)$ where $V$ and $E$ are respectively the set of vertices and edges, the MaxCut cost Hamiltonian reads:
\begin{equation}
    H_1 = \sum_{(i,j)\in E}w_{ij}\left(\frac{1-\sigma^z_i\sigma^z_j}{2}\right),
\end{equation}
where $w_{ij}\geq 0$ is the weight of edge $(i,j)$, which is set to 1 for unweighted graphs. Since the ground state of $-H_1$ encodes the optimal cut of $G$, we minimize the expectation value $-E(\psi(\vec{\theta}))$. Moreover, since we use non-negative weights and the MaxCut cost function to maximize is $C(x)=\sum_{(i,j)\in E}w_{ij}(x_i\oplus x_j)$, the worst cut value is $C_{\min}=0$. Flipping the sign of $H_1$ yields $E_{\min}\equiv C_{\max}$ with $C_{\max}$ the optimal cut value. Thus, the MaxCut approximation ratio is $r(\psi(\vec{\theta})) = E(\psi(\vec{\theta})) / E_{\min}$.

\begin{figure}[h!]
    \centering
    \includegraphics[scale=0.3]{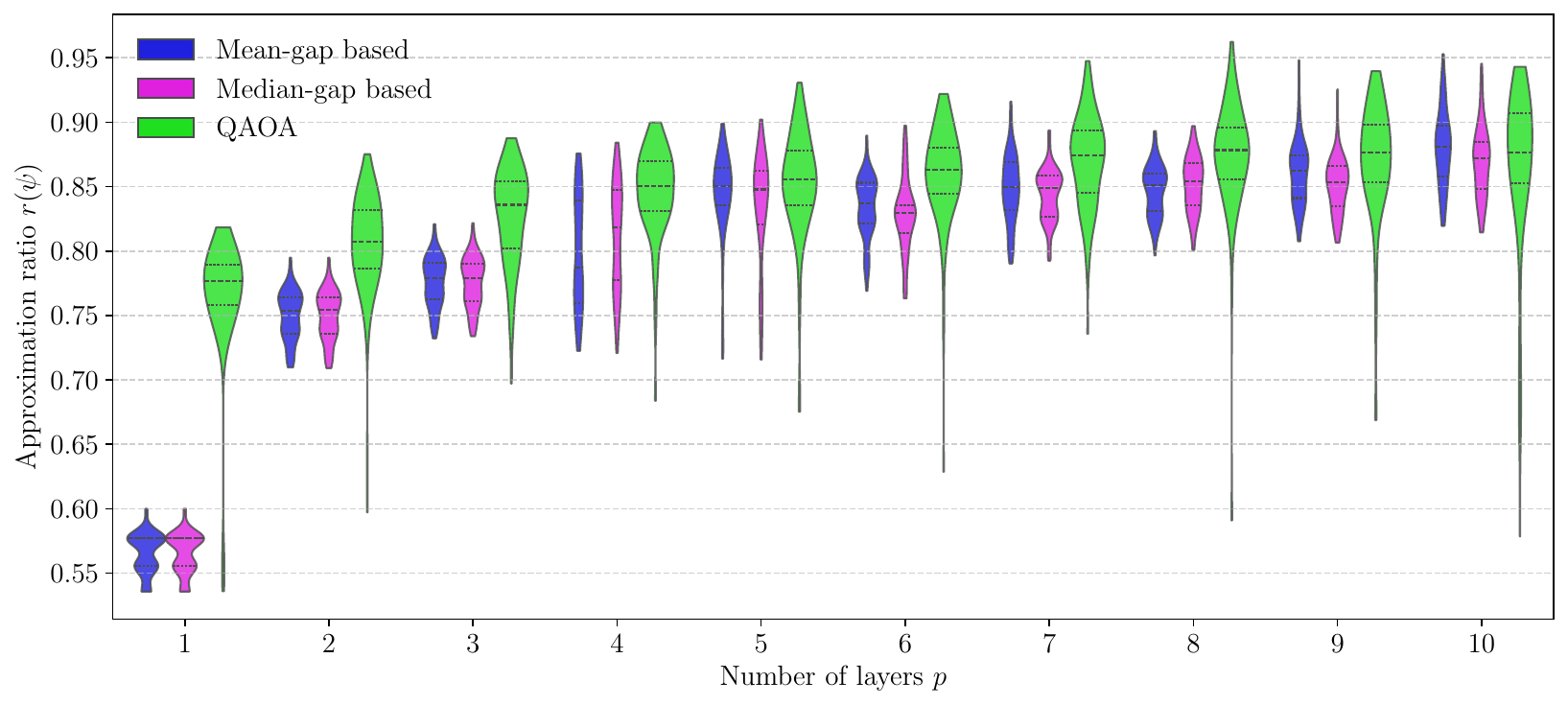}
    \includegraphics[scale=0.33]{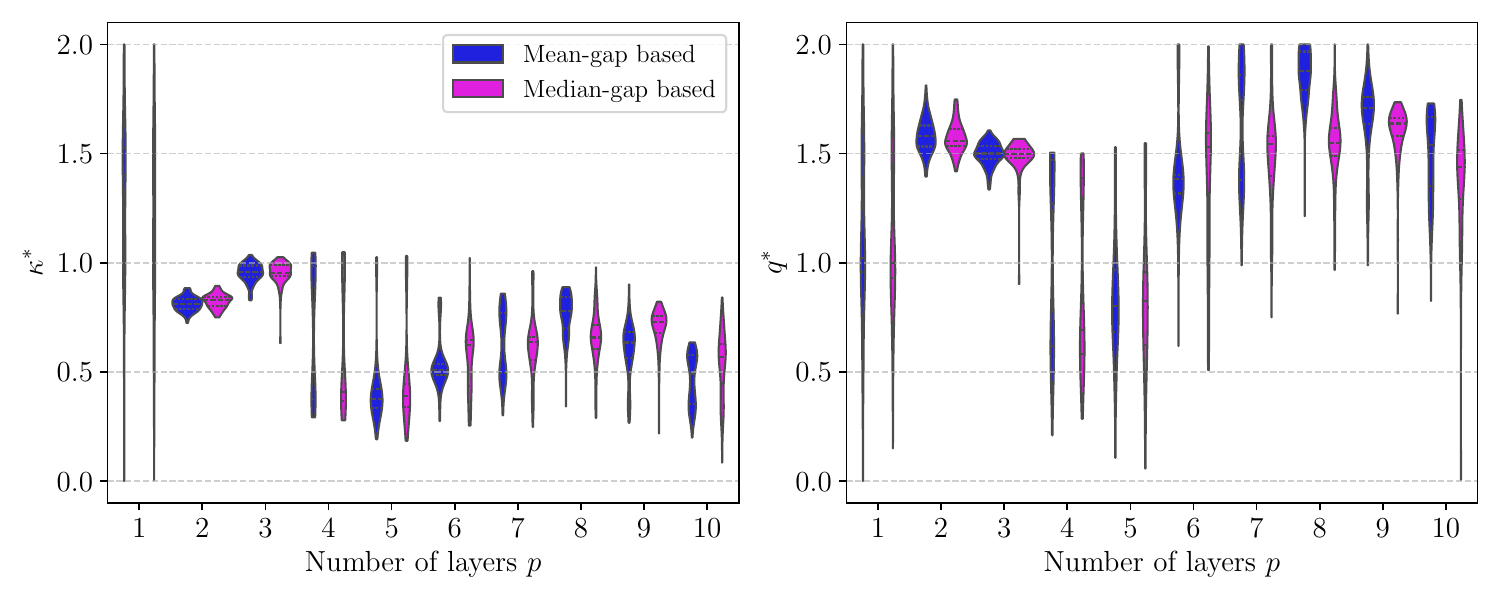}
    \caption{Experimental results of our heuristic and QAOA on 100 instances of 3-regular unweighted MaxCut problems of size $n=20$. Showing the approximation ratios (upper) and optimized hyperparameters obtained via classical optimization (lower). The learning part of our heuristic is done on random QUBOs of size $n=10$ and the classical optimization is done for $2$ (our heuristic) and $2p$ (QAOA) hyperparameters.}
    \label{fig:unweighted_maxcut_results}
\end{figure}

\begin{figure}[h!]
    \centering
    \includegraphics[scale=0.3]{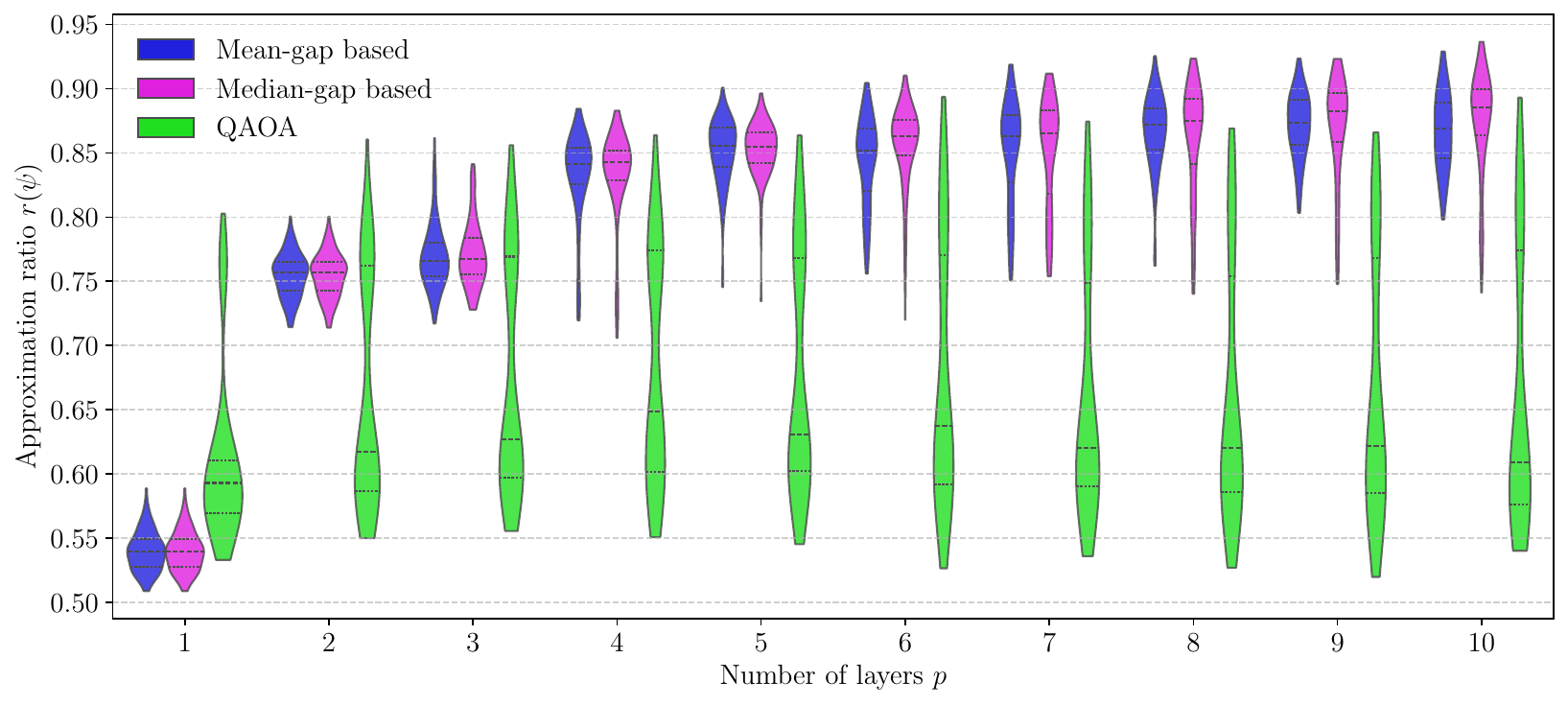}
    \includegraphics[scale=0.33]{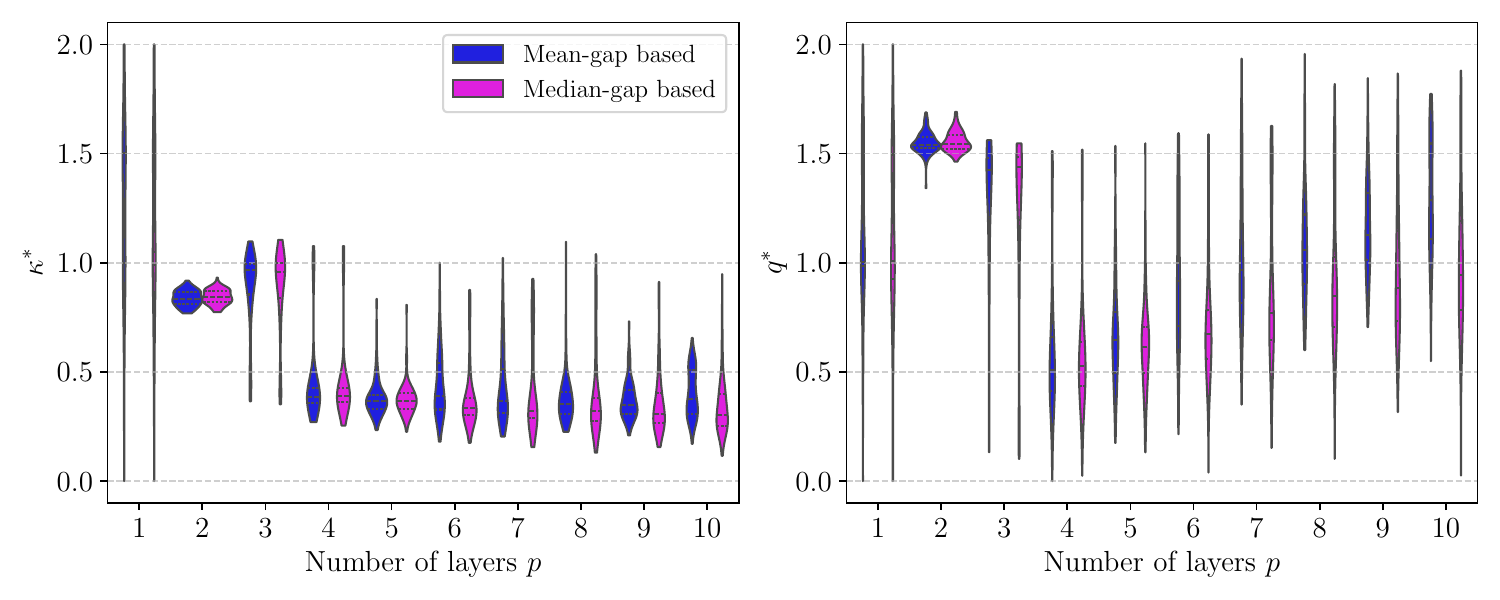}
    \caption{Experimental results of our heuristic and QAOA on 100 instances of 3-regular weighted MaxCut problems of size $n=20$. Showing the approximation ratios (upper) and optimized hyperparameters obtained via classical optimization (lower). The learning part of our heuristic is done on random QUBOs of size $n=10$ and the classical optimization is done for $2$ (our heuristic) and $2p$ (QAOA) hyperparameters.}
    \label{fig:weighted_maxcut_results}
\end{figure}

We generate 100 random unweighted and weighted ($w_{ij}\in[0,10]$) $3$-regular graphs of size $n=20$ and we compare the performance of our heuristic with QAOA. We show the results in Fig.~\ref{fig:unweighted_maxcut_results} and Fig.~\ref{fig:weighted_maxcut_results}. In the \(x\)-axis, the number of layers \(p\) represents the number of layers of the circuit on which we optimize, and does not represent mid-circuit measurement of the statevector. 

In both cases, the optimal parameters obtained via classical optimization display a similar depth dependence: angles are larger and more dispersed at shallow depth ($p\leq 3$), then decrease and progressively concentrate as $p$ increases. Theoretically, we expect $q$ to range in $(1,2)$, the numerical simulations here are not enough to confirm such behavior in practice. The parameter $\kappa$ acts as the inverse total runtime of the continuous evolution; at fixed $n$, we expect this to be constant. Moreover, we only observe minor differences between the mean-gap and median-gap heuristics, with nearly overlapping distributions as $p$ increases.

The corresponding approximation ratios improve steadily with depth for all methods. The gap-based heuristics achieve high-quality solutions already at moderate depth and continue to increase toward saturation at larger $p$, with the median-gap variant showing a slight but consistent advantage. Standard QAOA follows the same qualitative trend but remains more variable and, depending on the instance set (unweighted or weighted), generally does not surpass the gap-based strategies. Overall, these results indicate that gap-informed parameter choices yield robust performance gains while requiring progressively smaller optimal angles as the circuit depth grows.

\subsection{Random QUBOs}

\begin{figure}[h]
    \centering
    \includegraphics[scale=0.3]{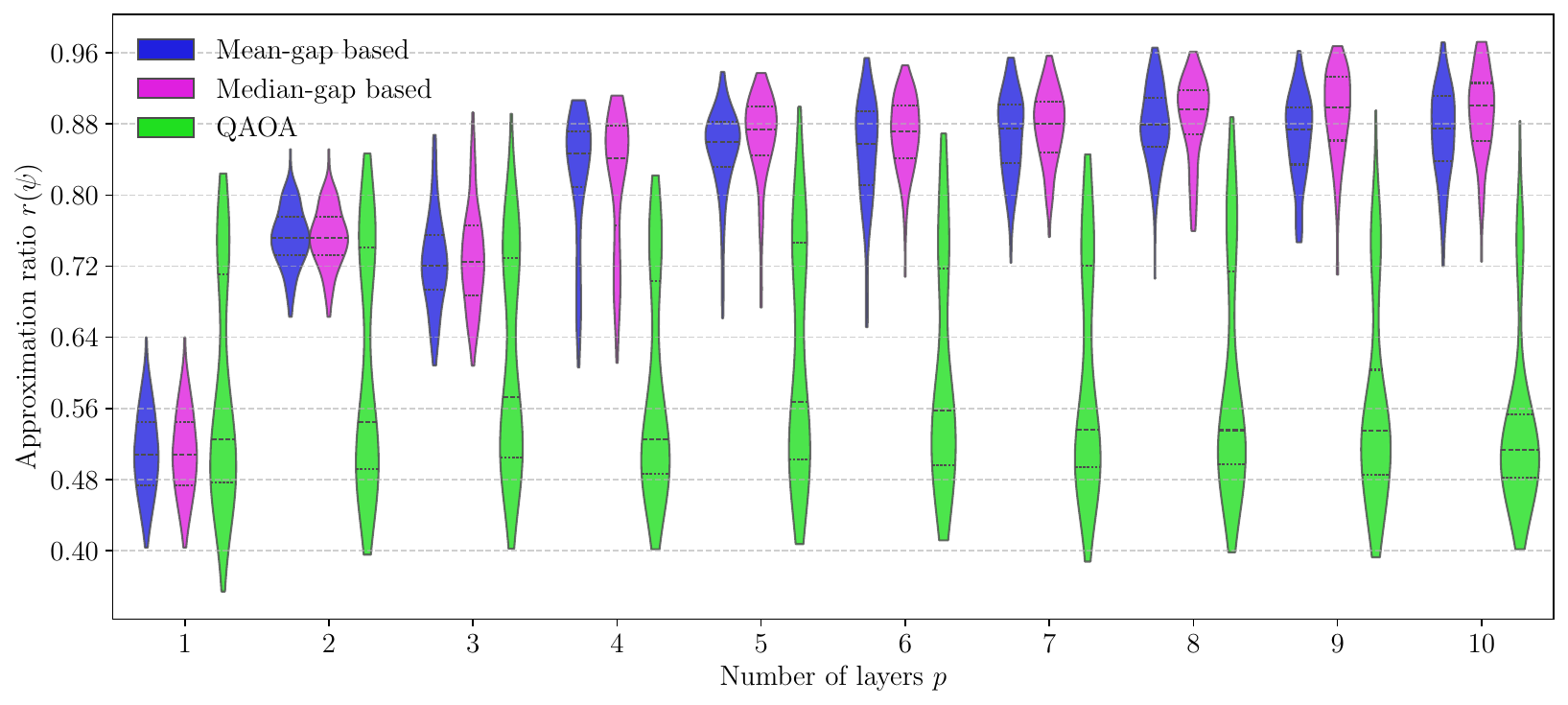}
    \includegraphics[scale=0.33]{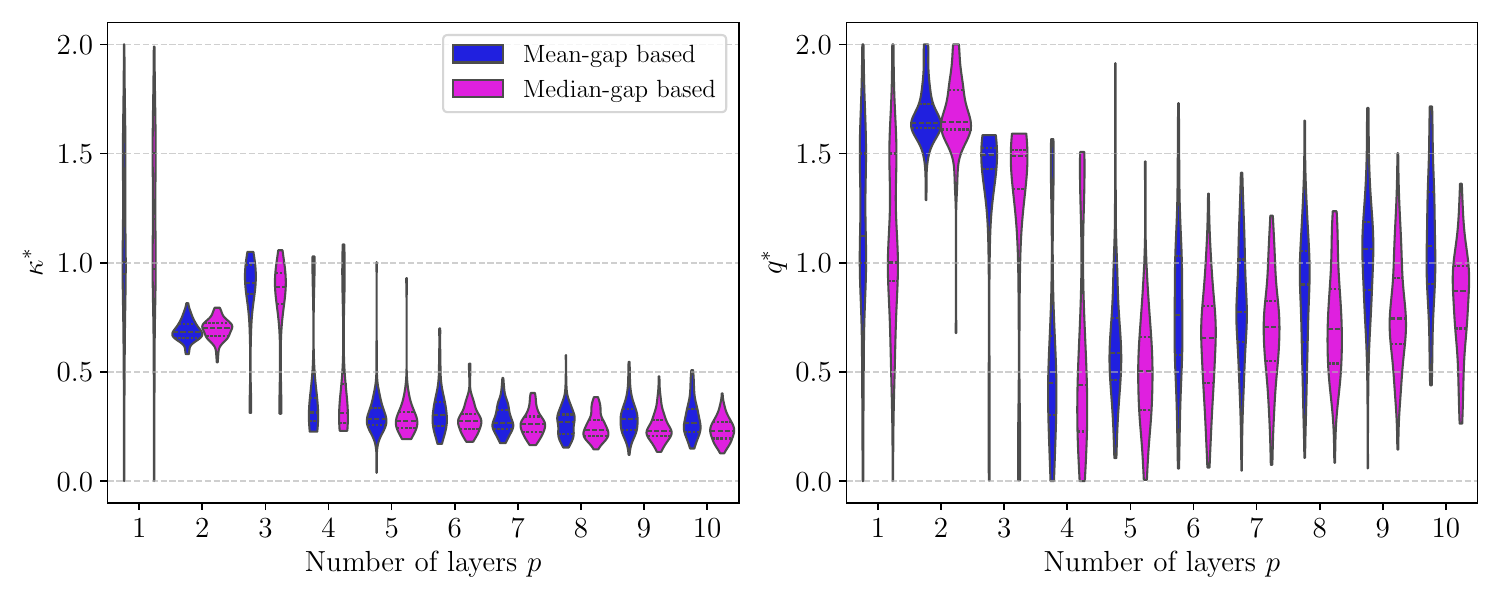}
    \caption{Experimental results of our heuristic and QAOA on 100 instances of random QUBO problems of size $n=20$. Showing the approximation ratios (upper) and optimized hyperparameters obtained via classical optimization (lower). The learning part of our heuristic is done on random QUBOs of size $n=10$ and the classical optimization is done for $2$ (our heuristic) and $2p$ (QAOA) hyperparameters.}
    \label{fig:random_results}
\end{figure}

We generate random QUBO problems with real coefficients $Q_{ij}$ uniformly drawn in $[-100,100]$. The cost Hamiltonian is that of Eq. \eqref{eq:ising_hamiltonian}, whose ground state encodes the optimal solution. We show the results in Fig. \ref{fig:random_results}.

Again, the two lower plots show the distributions of the optimal hyperparameters obtained via classical optimization for the mean-gap and median-gap heuristics. At low depth ($p \leq 3$), the hyperparameters are larger and more dispersed, indicating a sensitive optimization landscape. As the number of layers $p$ increases, they decrease and concentrate, stabilizing at small values for $p \geq 7$, with only minor differences between the two heuristics.

The higher plot reports the corresponding approximation ratios. Both gap-based heuristics consistently outperform standard QAOA and improve monotonically with depth, approaching high-quality solutions at larger $p$. The median-gap heuristic shows a slight advantage at higher depths. In contrast, standard QAOA displays lower ratios with large variance and little improvement as $p$ increases, highlighting the benefit of the gap-based parameter strategies.

\subsection{Performance difference with QAOA}

\begin{figure}[h]
    \centering
    \includegraphics[scale=0.45]{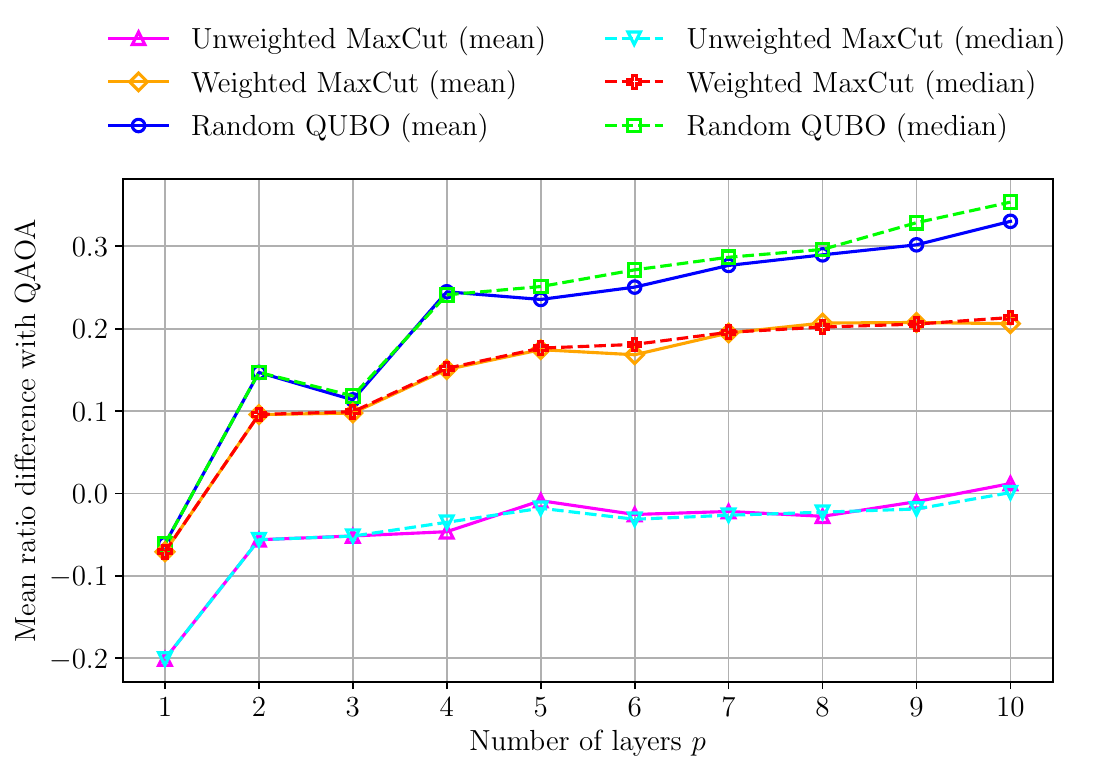}
    \caption{Mean approximation ratio difference between our heuristic and QAOA on unweighted 3-regular MaxCut, weighted 3-regular MaxCut and random QUBO problems of size $n=20$ as a function of the number of layers $p$.}
    \label{fig:difference_qaoa}
\end{figure}

Lastly, we compare the mean approximation ratio difference of our heuristic with vanilla QAOA in Fig. \ref{fig:difference_qaoa}. Overall, the gap-based strategies yield improvements over QAOA that grow with depth for weighted MaxCut and random QUBO, reaching the largest gains at high $p$. Moreover, random QUBO exhibits the strongest advantage, followed closely by weighted MaxCut. In contrast, the improvement for unweighted MaxCut remains modest and close to zero beyond shallow depth, indicating more limited benefit from gap-based parameter selection in this case. Furthermore, the mean and median based gap variants follow very similar trends, with only minor quantitative differences across all problem classes.

\section{Conclusion}

In this work, we introduced a spectral gap informed schedule learning framework for QAOA that transfers information from small-scale instances to larger systems. By mapping a continuous adiabatic schedule to its trotterized implementation, we obtain closed-form expressions for all QAOA angles in terms of only two global hyperparameters, eliminating the need to variationally optimize each layer. This construction reduces the classical optimization problem from \(2p\) independent parameters to a constant-dimensional problem, independent of circuit depth. 

Our numerical investigations on MaxCut and random QUBO instances indicate that the learnt schedules transfer effectively across system sizes, achieving better approximation ratios than QAOA for the most general kind of instances that are 3-regular weighted MaxCut and random QUBO, while significantly reducing classical optimization overhead. This could be explained by the fact that the dimension of the classical optimization problem is significantly lower for our method. Moreover, the reduced parameter space naturally mitigates sensitivity to barren plateau phenomena, offering a more stable training procedure for deep circuits. 

More broadly, our results provide some evidence that structural information extracted from small instances, here in the form of spectral gap profiles, can serve as a principled inductive scheme for variational quantum algorithms. 

\addtolength{\textheight}{-3cm}   


\section{Acknowledgments}

The authors thank Andrea Simonetto for his useful feedback on
the form and content of this manuscript. This research benefited from the support of the FMJH Program PGMO, project number 2023-0009, the ANR project HQI-ANR-22-PNCQ-0002, the PEPR EPiQ ANR-22-PETQ-0007, the ANR JCJC DisQC ANR-22-CE47-0002-01. AB was supported in part by the French PEPR integrated project HQI (ANR-22-PNCQ-0002) and the French ANR project QUOPS (ANR-22-CE47-0003-01).


\bibliographystyle{unsrt} 
\bibliography{ref} 






\end{document}